# Spectroscopic observation of SU(*N*)-symmetric interactions in Sr orbital magnetism

X. Zhang[1,2], M. Bishof[1,2], S. L. Bromley[1,2], C.V. Kraus[3,4], M. S. Safronova[5,6], P. Zoller[3,4], A. M. Rey[1,2], J. Ye[1,2]

[1]*JILA, National Institute of Standards and Technology and University of Colorado, Boulder, CO 80309-0440, USA*
[2]*Department of Physics, University of Colorado, Boulder, CO 80309-0390, USA*
[3]*Institute for Quantum Optics and Quantum Information of the Austrian Academy of Sciences, A-6020 Innsbruck, Austria*
[4]*Institute for Theoretical Physics, University of Innsbruck, A-6020 Innsbruck, Austria*
[5]*Department of Physics and Astronomy, University of Delaware, Newark, DE 19716, USA*
[6]*Joint Quantum Institute, National Institute of Standards and Technology and the University of Maryland, College Park, Maryland, 20742, USA*

SU(*N*) symmetry can emerge in a quantum system with *N* single-particle spin states when spin is decoupled from inter-particle interactions. So far, only indirect evidence for this symmetry exists, and the scattering parameters remain largely unknown. Here we report the first spectroscopic observation of SU(*N*=10) symmetry in $^{87}$Sr using the state-of-the-art measurement precision offered by an ultra-stable laser. By encoding the electronic orbital degree of freedom in two clock states, while keeping the system open to 10 nuclear spin sublevels, we probe the non-equilibrium two-orbital SU(*N*) magnetism via Ramsey spectroscopy of atoms confined in an array of two-dimensional optical traps. We study the spin-orbital quantum dynamics and determine all relevant interaction parameters. This work prepares for using alkaline-earth atoms as test-beds for iconic orbital models.

**Keywords:** SU(*N*), Orbital, Magnetism

Symmetries play a fundamental role in the laws of nature. A very prominent example is SU(*N*) symmetry as the source of intriguing features of quantum systems. For instance, the SU(3) symmetry of quantum chromodynamics governs the behavior of quarks and gluons. When generalized to large *N*, it is anticipated to give rise to a large degeneracy and exotic many-body behaviors. Owing to the strong decoupling between the electronic-orbital and nuclear-spin degrees of freedom [1], alkaline-earth (-like) atoms, prepared in the two lowest electronic states (clock states), are predicted to obey SU(*N*=2*I*+1) symmetry with respect to the nuclear spin (*I*) [2–5]. Thanks to this symmetry, in addition to their use as ideal time keepers [6] and quantum information processors [7], alkaline earth atoms are emerging as a unique platform for the investigation of high-energy lattice gauge theories [8], for testing iconic orbital models used to describe transition metal oxides, heavy fermion compounds, and spin liquid phases [9], and for the observation of exotic topological phases [5,10]. Progress towards these goals includes the production of quantum degenerate gases for calcium [11] and all stable isotopes of strontium and ytterbium [12,13], the capability of imaging individual spin components via optical Stern-Gerlach methods [14], and control of interactions with optical Feshbach resonances [12,15,16]. Furthermore, the best atomic clock has been produced with lattice confined Sr atoms [6], and many-body spin dynamics have been studied directly in that system [17].

However, thus far only indirect evidence for SU(*N*) symmetry exists, including inference from suppressed nuclear spin-relaxation rates [14], reduced temperatures in a Mott insulator for increased number of spin states [18], and the changing character of a strongly-interacting one-dimensional fermionic system as a function of *N* [19]. Furthermore, these observations are limited to the electronic ground state. The corresponding ground-state *s*-wave scattering parameter, $a_{gg}$, has been determined from photo-association [20] and rovibrational spectroscopy [21], but the excited state-related scattering parameters remain unknown.

In this paper, we report the first spectroscopic observation of SU(*N*) symmetry and two-orbital SU(*N*) magnetism in an ensemble of fermionic $^{87}$Sr atoms at μK temperatures and confined in an array of two-dimensional (2D) disc-shaped, state-insensitive optical traps [22]. The axial (Z) trapping frequency $\nu_Z$ is ~80 kHz and the radial (X-Y) frequency $\nu_R$ is ~600 Hz. The SU(*N*) symmetric spin degree of freedom is encoded in the 10 nuclear spin states with quantum number $m_I$ (Fig. 1A), and the pseudo-spin ½ orbital degree of freedom in the two lowest electronic

(clock) states ($^1S_0$ and $^3P_0$, henceforth $|g\rangle$ and $|e\rangle$). Under typical atomic occupancies (< 20 atoms per disc), temperatures (1 µK< $T_R$ <7 µK, $T_Z$ ~2 µK), and trap volume, the mean interaction energy per particle is at least two orders of magnitude smaller than the single-particle vibrational spacing along any direction. The unprecedented spectral resolution available with an ultra-stable laser of 1 × $10^{-16}$ stability [23] enables us to accurately probe these interactions while addressing individual nuclear spin levels.

To the first order approximation, atoms are frozen in the initially populated motional energy modes and the quantum dynamics takes place only in the internal degrees of freedom (spin and orbital) [17,24]. Atoms distributed among these quantized motional levels are thus analogous to atoms localized in real-space lattice trapping potentials. Moreover, the *s*-wave and *p*-wave (Fig. 1B) interactions, which generate the dynamics, couple the atoms without being overly sensitive to the motional eigenenergies, thus providing nonlocal interactions when viewed within a lattice spanned by eigenenergies. This allows us to study spin lattice models with effective long-range couplings in a non-degenerate Fermi gas (Fig. 1C). Spin models with long-range interactions have been implemented in dipolar gases [25] or trapped ionic systems [26], but our system has the additional SU(*N*) symmetry to enrich the manybody dynamics. By performing Ramsey spectroscopy with various nuclear spin mixtures, we determine the nuclear spin independence of the *s*-wave and *p*-wave interactions. Furthermore, we probe the non-equilibrium dynamics of the orbital coherence, and the results are well reproduced by a two-orbital SU(*N*) spin lattice model in quantized motional eigenenergy space.

Interactions between two $^{87}$Sr atoms are governed by Fermi statistics with an overall antisymmetrization under exchange in the motional, electronic, and nuclear spin degrees of freedom (Fig. 1B). Consider a pair of interacting atoms (*j* and *k*) occupying two of the quantized eigenmodes of the trapping potential, $\mathbf{n}_j$ and $\mathbf{n}_k$. If the atoms are in a nuclear spin symmetric state they experience *s*-wave interactions only if their electronic state is anti-symmetric: $(|eg\rangle - |ge\rangle)/\sqrt{2}$. We denote the elastic scattering length characterizing those collisions as $a_{eg}^-$. They can collide via *p*-wave interactions in three possible electronic symmetric configurations $\{|gg\rangle, |ee\rangle, (|eg\rangle + |ge\rangle)/\sqrt{2}\}$, corresponding to the *p*-wave elastic scattering lengths $b_{gg}, b_{ee}, b_{eg}^+$, respectively. In contrast, if the two atoms are in an anti-symmetric nuclear spin configuration they experience *s*-wave collisions under these three electronic symmetric configurations, with the corresponding scattering lengths $a_{gg}, a_{ee}, a_{eg}^+$, respectively. Accordingly, *p*-wave interactions occur in $(|eg\rangle - |ge\rangle)/\sqrt{2}$, corresponding to the scattering length $b_{eg}^-$. These eight parameters characterize elastic collisions at ultralow temperatures, and SU(*N*) symmetry predicts them to be independent of the nuclear spin configuration. Here, *N* is chosen by initial state preparation and can vary from 1 to 10 in $^{87}$Sr (*I*=9/2). The Hamiltonian that governs these interactions can be written in terms of orbital-spin 1/2 operators $\hat{T}_j^{x,y,z}$ acting on the *j*-atom's electronic state, $\{e,g\}$, and in terms of nuclear-spin permutation operators $\hat{S}_n^m(j)$, acting on the *j*-atom's nuclear spin levels, *n, m* $\in \{1,2,\ldots N\}$ as:

$$\hat{H} = (\hat{\mathcal{P}}^+\hat{H}^+ + \hat{\mathcal{P}}^-\hat{H}^-) \ , \quad (1) \text{ and}$$

$$\hat{H}^\pm = J_{j,k}^\pm \vec{T}_j \cdot \vec{T}_k + \chi_{j,k}^\pm \hat{T}_j^z \hat{T}_k^z + C_{j,k}^\pm \left(\frac{\hat{T}_j^z+\hat{T}_k^z}{2}\right) + K_{j,k}^\pm \hat{\mathbb{I}}. \quad (2)$$

Here, $\hat{\mathbb{I}}$ is the identity matrix, $\hat{\mathcal{P}}^\pm = \frac{[\hat{\mathbb{I}} \pm \sum_{\alpha,\beta} \hat{S}_\beta^\alpha(j)\hat{S}_\alpha^\beta(k)]}{2}$ are nuclear spin projector operators into the symmetric triplets (+) and anti-symmetric singlet (-) nuclear spin states, respectively. Eq. (1) states that if the nuclear spin of the atoms is in (+) or (-), then they interact according to $\hat{H}^+$ or $\hat{H}^-$, respectively. The coupling constants $J_{j,k}^\pm, \chi_{j,k}^\pm, C_{j,k}^\pm, K_{j,k}^\pm$ depend on the scattering parameters, $a_\eta$ and $b_\eta$, $\eta \in \{ee, gg, eg^+ \text{ and } eg^-\}$, and the wavefunction overlap of the *j* and *k*-atom's vibrational modes (See Supplementary Material). The Hamiltonian commutes with all the SU(*N*) generators, $\hat{S}_n^m(j)$, and is thus invariant under transformations from the SU(*N*) group (i.e., SU(*N*) symmetric). In addition to elastic interactions, $^{87}$Sr atoms also exhibit inelastic collisions. Among those however, only the *e-e* ones have been observed to give rise to measureable losses [27]; we denote these two inelastic scattering lengths as $\gamma_{ee}$ and $\beta_{ee}$ for *s*-wave and *p*-wave, respectively. We set other inelastic parameters to zero based on their negligible contributions in measurements.

We first test SU(*N*) symmetry in a two-orbital system by measuring the density-dependent frequency shift of the clock transition under various nuclear spin population distributions. We use a Ramsey sequence to measure interactions [17] under an external magnetic field that produces Zeeman splittings much larger than the interaction energy. As shown in Fig. 2A, the sequence starts with all atoms in $|g\rangle$. Only atoms in a particular nuclear spin state are coherently excited and interrogated, while atoms in other states ("spectators") remain in $|g\rangle$. We denote $\mathcal{N}_i^{\text{tot}}$ the number of interrogated atoms, $\mathcal{N}_S^{\text{tot}}$ the number of spectator atoms, and define a population ratio $f = \mathcal{N}_S^{\text{tot}}/\mathcal{N}_i^{\text{tot}}$ and the interrogated fraction $x_i = \mathcal{N}_i^{\text{tot}}/(\mathcal{N}_i^{\text{tot}} + \mathcal{N}_S^{\text{tot}})$. We control orbital excitation, $p_e$, by varying the initial pulse area, $\theta_1$, in $0 < \theta_1 < \pi$. After a free evolution time, $\tau_{\text{free}} = 80$ ms, a second pulse of area $\pi/2$ is applied for subsequent readout. The resonance frequency shift is recorded as the atomic number in the trap is varied. We operate with highly homogeneous atom-laser coupling. Consequently, the *p*-wave interaction in a fully spin-polarized sample is dominant [17].

In Fig. 2B, we compare the fully spin-polarized case ($m_I$ = +9/2) against three other scenarios with different spin mixtures under $T_R$=6-7 µK. The observed density shifts as a linear function of $p_e$, when scaled to the same number of interrogated atoms ($\mathcal{N}_i^{\text{tot}}$ =4000), show three features: **(I)** the linear slope, *l*, depends only on $\mathcal{N}_i^{\text{tot}}$, **(II)** the offset with respect to the polarized case linearly increases with *f*, and **(III)** both *l* and the offset are independent of how the atoms are distributed in the nuclear spin levels. The latter is verified by measuring the same shifts when interrogating 29% of the total population in either +9/2 or +7/2.

To determine the temperature dependence for the density shift and for additional confirmation of the observed nuclear spin independence, we interrogate other nuclear spin states, -9/2 or -3/2, under a lower $T_R$ ~2 μK, when the distribution across all spin states is nearly even (Fig. 2C). The measured density shifts scaled to $\mathcal{N}_i^{\text{tot}} = 4000$ are again similar to each other, providing further direct experimental evidences for SU(*N*=10) symmetry. At this lower $T_R$, while the slope depends only on $\mathcal{N}_i^{\text{tot}}$, there is a smaller offset of the density shift relative to the polarized case when $x_i$ varies. To quantify the $T_R$ dependence, we plot together all measured ratios, $l/l_0$, with $l_0$ the linear slope for the polarized case. We see that **(IV)** the ratios collapse into a single value independently of *f* and $T_R$ for fixed $\mathcal{N}_i^{\text{tot}}$, yielding $l/l_0 = 1.00\pm0.03$ (Fig. 2D). This result agrees well with the SU(*N*)-predicted ratio of unity and verifies this symmetry to the 3% level. We observe that *l* decreases only by 10% when $T_R$ is raised from 2 μK to 6 μK, verifying its insensitivity to $T_R$. We also determine the excitation fraction where the shift is zero in a spin mixture, $p_e^*$, and compare it to that of a polarized sample, $p_{e0}^*$ (gray bands, Fig. 2C), under various interrogated spin states (colors in Fig. 2E). The difference shows the following features: **(V)** it collapses onto a single line (for a given $T_R$ of either 2.3 or 6.5 μK) as a function of *f*, which provides a further evidence for spin-independence of the interactions; **(VI)** at $\mathcal{N}_s^{\text{tot}} = 0$ (fully polarized), the two lines cross each other at the origin, as expected from the $T_R$-insensitivity of the *p*-wave interactions. For $f > 0$, the proportionality constant is finite for 6.5 μK (lower line), and decreases to almost zero for $T_R$~2.3 μK (upper line). This near zero proportionality constant for $T_R$~2.3 μK indicates an accidental cancellation of the spectators' *s*- and *p*-wave interaction effects at this temperature.

In the presence of a large external magnetic field that produces differential Zeeman splittings much larger than the interaction energy, those terms in the Hamiltonian that exchange the population between the occupied spin-orbital levels are energetically suppressed and the populations of different spin-orbital levels are conserved. Hence, the Hamiltonian is dominated by Ising-type interactions that preserve the spin-orbital population. In this regime the many-body dynamics for a single trap with $\mathcal{N}$ atoms can be captured under a collective approximation that replaces the coupling constants with their corresponding thermal averages, $O_{n_j,n_k}^{\pm} \to \overline{O^{\pm}}$ [24]. For the experimental relevant case where only $\mathcal{N}_i$ atoms in spin $m_I$ are interrogated and where $\mathcal{N}_S$ atoms in the other spin components remain in $|g\rangle$, the effective many-body Hamiltonian during $\tau_{\text{free}}$ simplifies substantially. It consists of two parts, $\hat{H}_i + \hat{H}_S$. The first part, $\hat{H}_i = \overline{\chi^+}(\hat{\mathbb{T}}^z)^2 + \overline{C^+}\hat{\mathbb{T}}^z \mathcal{N}_i$, describes the *p*-wave interactions between the interrogated atoms [17,24], where $\hat{\mathbb{T}}^{\alpha=x,y,z} = \sum_j^{\mathcal{N}} \hat{S}_{m_I}^{m_I}(j) \hat{T}_j^{\alpha=x,y,z}$ are collective orbital operators acting on the $\mathcal{N}_i$ interrogated atoms. The density shift induced by these interactions, $\Delta v^i = \mathcal{N}_i (\overline{C^+} - \cos\theta_1 \overline{\chi^+})$, with $\overline{\chi^+} = \frac{[b_{ee}^3 + b_{gg}^3 - 2b_{eg}^{+3}]}{2}\langle P\rangle_{T_R}$ and $\overline{C^+} = \frac{[b_{ee}^3 - b_{gg}^3]}{2}\langle P\rangle_{T_R}$, depends linearly on the number of excited atoms $\mathcal{N}_i p_e$. Here $\langle P\rangle_{T_R}$ corresponds to the thermal average of the *p*-wave mode overlap coefficients. Assuming a Boltzmann distribution of initially populated radial motional modes, we have $\langle P\rangle_{T_R} \propto (T_R)^0$ (insensitive to $T_R$) [17]. For a spin polarized sample, the observed density shifts are well reproduced by theory (solid black lines in Figs. 2C) based on the same *p*-wave parameters as determined in Ref. 17. The second part, $\hat{H}_S = \mathcal{N}_S \overline{\Lambda} \hat{\mathbb{T}}^z$, describes the interactions between the interrogated and spectator atoms with both *p*- and *s*-wave contributions. The related density shift is $\Delta v^S = \overline{\Lambda} \mathcal{N}_S$, with
$$\overline{\Lambda} = \frac{\overline{C^+} + \overline{C^-} - \overline{J^+} - \overline{J^-} - \overline{\chi^+} - \overline{\chi^-}}{2} = \frac{(a_{eg}^+ + a_{eg}^- - 2a_{gg})}{4}\langle S\rangle_{T_R} + \frac{[b_{eg}^{+3} + b_{eg}^{-3} - 2 b_{gg}^3]}{4}\langle P\rangle_{T_R}.$$
The *s*-wave thermal average, $\langle S\rangle_{T_R}$, decreases with $T_R$ as $\langle S\rangle_{T_R} \propto \frac{1}{T_R}$.

This model fully reproduces the experimental observations as listed in **(I-VI)** and shown in Fig. 2. To quantitatively compare with the experiment, we perform a Poissonian average of the atom number across the 2D traps and use the average excitation fraction to account for the two-body *e-e* losses [17,27] during the free evolution. The capability of the SU(*N*) spin lattice model to reproduce the experimental observations also enables us to determine the remaining *s*- and *p*-wave scattering parameters. For each of the four channels, $\eta \in \{ee, gg, eg^+ \text{ and } eg^-\}$, the *s*-wave and *p*-wave parameters relate to each other through the characteristic length, $\bar{a}_\eta$, of the van der Waals potential [28]. Thus, after we determine $\bar{a}_\eta$ using the available van der Waals $C_6$ coefficients (see Supplementary Materials), only four elastic scattering parameters remain independent. Among those, $a_{gg}, b_{ee}, b_{eg}^+$ (and thus their respective *p*- or *s*-wave counterparts) are known [17], leaving only one unknown parameter associated with the *eg*- channel. We fit the data in Fig. 2E and extract $a_{eg}^-$ and $b_{eg}^-$. Table 1 lists all the scattering parameters determined from the prior and current measurements. We emphasize that the test of SU(*N*) symmetry (at the 3% level) is based directly on the measured interactions that are independent of nuclear spin configurations, and it does not require accurate knowledge of some common-mode system calibrations.

To reveal the SU(*N*) orbital magnetism in our system we need to perform coherent dynamic spectroscopy. Ramsey spectroscopy is particularly suitable for this goal, as the atomic evolution during the field-free period gives rise to variable $e$-$g$ orbital coherence. We measure the decay of the $e$-$g$ coherence, in the form of Ramsey fringe contrast $\mathcal{C}(\tau_{\text{free}}) = 2/\mathcal{N}_i^{\text{tot}} \sqrt{\langle \hat{\mathbb{T}}_{\text{tot}}^x \rangle^2 + \langle \hat{\mathbb{T}}_{\text{tot}}^y \rangle^2}$ as a function of $\tau_{\text{free}}$, under various population distributions among nuclear spin states (Fig. 3A). Here $\hat{\mathbb{T}}_{\text{tot}}^{x,y}$ is the sum of $\hat{\mathbb{T}}^{x,y}$ over the 2D traps. In the presence of a large magnetic field, the decay of $\mathcal{C}$ has two sources. The first arises from within the interrogated atoms: $p$-wave elastic interactions, two-body $e$-$e$ losses, higher-order interaction-induced mode-changing processes, as well as dephasing induced by the distribution of atoms across traps. All these $p$-wave effects are accounted for in our theory using the same $p$-wave parameters [17]. The second source comes from spectators, which act on the interrogated atoms at a given site as an inhomogeneous and density-dependent effective magnetic field along $z$, with both $s$- and $p$-wave contributions. The effective magnetic field is static if the atoms are frozen in their motional states, but can vary with time in the presence of higher-order mode-changing processes. The $p$-wave interaction plays a dominant role at high $T_R$ = 5-6 μK, while the $s$-wave interaction, which has a stronger dependence on mode distribution, becomes significant at lower $T_R$.

To understand the orbital dynamics in detail, we first study three different cases under $T_R$ = 5-6 μK, with $\theta_1 = \pi/4$ and the interrogated fraction $x_i = 100\%$, 14%, and 56%, as displayed in Figs. 3B, C, E. To separate the effects of dephasing and many-body correlation in the contrast decay, we apply a $\pi$ echo pulse in the middle of the Ramsey sequence (Fig. 3A, lower panel). At the end of the sequence, the dephasing caused by the interactions between interrogated and spectator atoms and other technical effects is removed by the echo. The $\pi$ echo pulse modifies the contrast decay in a $\theta_1$-dependent way, because the $p$-wave contribution to contrast decay contains both $\theta_1$-dependent and -independent terms, as well as enhanced $e$-$e$ loss after the echo pulse for $\theta_1 < \pi/2$. The $\theta_1$-dependent contribution is generated by the term $\overline{\chi^+}(\hat{\mathbb{T}}^z)^2$ in the Hamiltonian, and can lead to many body orbital correlations that are unremovable by echo. The $\theta_1$-independent contribution is generated by the term $\overline{C^+}\hat{\mathbb{T}}^z \mathcal{N}_i$, and can be removed by a $\pi$ echo.

As shown in Fig. 3B, in a polarized sample and under $\theta_1 = \pi/4$, the Ramsey contrast decays more slowly with an echo pulse. This positive echo effect can be attributed to the suppressed dephasing from inhomogeneous atomic densities across different 2D traps ($\theta_1$-independent contribution) and to the faster number loss (note the number normalization in $\mathcal{C}$) with echo. Figures 3C and 3E show similar positive effect of an echo pulse in the presence of spectator atoms. Since $p$-wave interactions between interrogated atoms are reduced as the interrogated fraction decreases, the overall contrast decay becomes slower. Based on the determined scattering parameters, our model predicts that spectator atoms cause almost negligible decoherence effects at this high $T_R$ = 5-6 μK (see Supplementary Material).

In a polarized sample where $p$-wave interactions dominate, the contrast decay is expected to be insensitive to $T_R$. This is confirmed in Fig. 3D where measurements at $T_R$ = 2.6 μK show similar decay behaviors to those at 5.4 μK (Fig. 3B). Figure 3F plots the ratio of contrasts with and without echo and illustrates the positive echo effect in suppressing contrast decay in a fully polarized sample under $\theta_1 = \pi/4$, as well as the negative effect under $\theta_1 = 3\pi/4$ when the echo enhances contrast decay. The negative echo effect can be attributed to both the development of many-body orbital correlations under $\theta_1 = 3\pi/4$ [17] and the reduced $e$-$e$ loss after the echo, and is well reproduced by our spin lattice model.

When we lower $T_R$ to ~2 μK, the rise of the $s$-wave contribution causes significant decoherence effects due to the spectator atoms. Figure 4A illustrates the $x_i = 14\%$ minority case where contrast decay is clearly faster than in Fig. 3C, showing the influence of spectators. The inclusion of off-resonant mode-changing collisions as higher order corrections is now required to accurately reproduce the experimental observations (see Supplementary Materials). These mode-changing collisions can be visualized as relocating pairs of atoms in the energy-space lattice shown in Fig. 1C, analogous to interaction-induced tunneling processes in a real space lattice. The echo pulse suppresses the part of contrast decay arising from mode-preserving collisions between spectators and interrogated atoms, but it cannot reverse the decay due to mode-changing processes. In Fig. 4A, the measured contrast decay with echo enables us to determine a single parameter characterizing the mode-changing processes (see Supplementary Materials).

For a further and independent test of our model, we explore another case with $x_i = 56\%$ and $T_R$ ~2 μK, so that both the interrogated atoms and spectator atoms have important contributions to the contrast decay. As shown in Figs. 4B, D, the data are well described by the same theory model without varying any parameters, demonstrating a firm understanding of the system dynamics.

The experimental exploration of exotic SU($N$) physics is just starting. The unique capability of precision laser spectroscopy has so far allowed us to explore Ising orbital magnetism at relatively high temperatures. We expect to explore the full Hamiltonian including the exchange interactions by controlling the atomic density, temperature and the magnetic field to engineer various spin-spin and spin-orbital dynamics. This will allow us to push the frontier of emergent many-body quantum physics at increasingly high temperatures, as well as the study of time-resolved dynamics in the SU($N$) Kondo lattice and Kugel-Khomskii models [4,29,30] in the quantum gas regime.

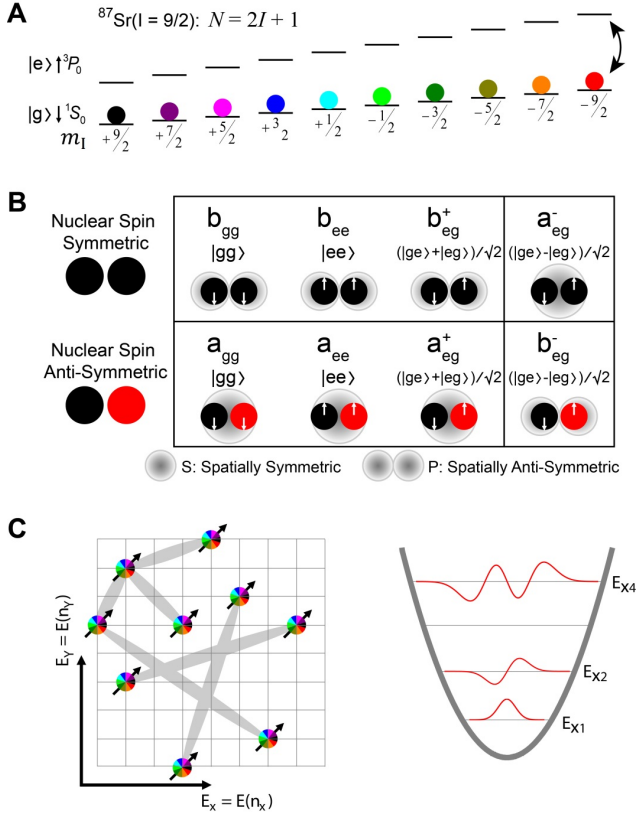
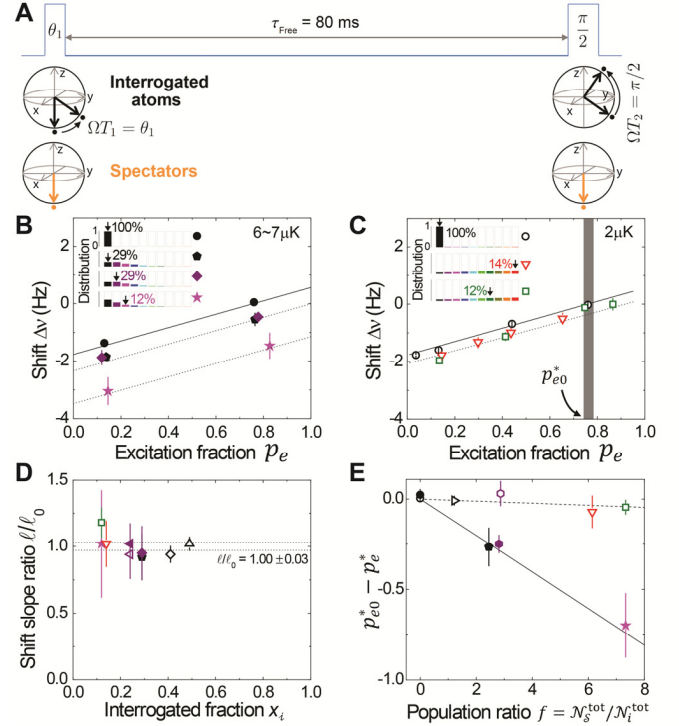

**Fig. 1. Diagram of the interacting spin lattice.** (**A**) Energy levels for the two lowest electronic states ($^1S_0$ and $^3P_0$) of $^{87}$Sr atoms in a magnetic field, each with ten nuclear spin states, depicted by colors. This color scheme is used throughout the paper to denote the interrogated state. (**B**) Interactions between two fermionic atoms characterized by four s-wave ("$a$") and four p-wave ("$b$") elastic scattering parameters. The interactions are governed by symmetries in motional states (bottom labels), nuclear spins (left labels), and electronic orbitals (white arrows). (**C**) (Left) Interacting electronic orbitals (spin-½ arrows) distributed over a lattice spanned by motional eigen-energies. Colored circles show the possibility of preparing coherent superpositions or statistical mixtures of $N$ nuclear spin states. (Right) Illustration of a few lowest occupied eigenmodes of a harmonic trap.

**Fig. 2. Nuclear spin independence of interaction effects.** (**A**) Ramsey sequence with an initial pulse of area $\theta_1$, a final $\pi/2$ pulse, and a free evolution period $\tau_{\text{free}} = 80$ ms. The spectator atoms remain in $|g\rangle$. (**B**) and (**C**) Measured density shifts (in symbols) for different nuclear spin configurations at $T_R = 6$-$7$ μK and $\sim 2$ μK, respectively. For consistency, the shifts are scaled for $\mathcal{N}_i^{\text{tot}} = 4000$. The inset illustrates the interrogated states (black arrows) and population distributions among various nuclear spin states. Solid and dotted lines show theory calculations for the corresponding $x_i$ and $T_R$ as indicated in the plots. The gray band in (**C**) corresponds to $p_{e0}^*$, the excitation fraction for zero density shift in a polarized sample. The spectator atoms generate a temperature-dependent density shift, which is independent of $p_e$ of the interrogated atoms and thus manifests as a net offset from the purely polarized density shift. (**D**) Ratio of the slope of the frequency shift between the spin mixed and polarized samples. The dotted lines represent the standard error. (**E**) The difference in the zero-shift excitation fraction between the spin mixed and polarized samples. The solid and dashed theory lines are used to determine the $b_{eg}^-$ and $a_{eg}^-$ scattering parameters. In **D** and **E**, two values of $T_R$ are used: 2.3(2) μK (open symbols) and 6.5(4) μK (filled symbols). In addition to conditions used for **B** and **C**, other spin configurations are studied: open up triangles ($x_i = 49\%$), open diamond (41%), open right triangles (46%), open and filled hexagons (26%), open and filled left triangles (24%), filled pentagons (29%), and filled stars (12%). Error bars represent 1σ standard error inflated by the square root of the reduced chi-squared, $\sqrt{\chi_{\text{reduced}}^2}$.

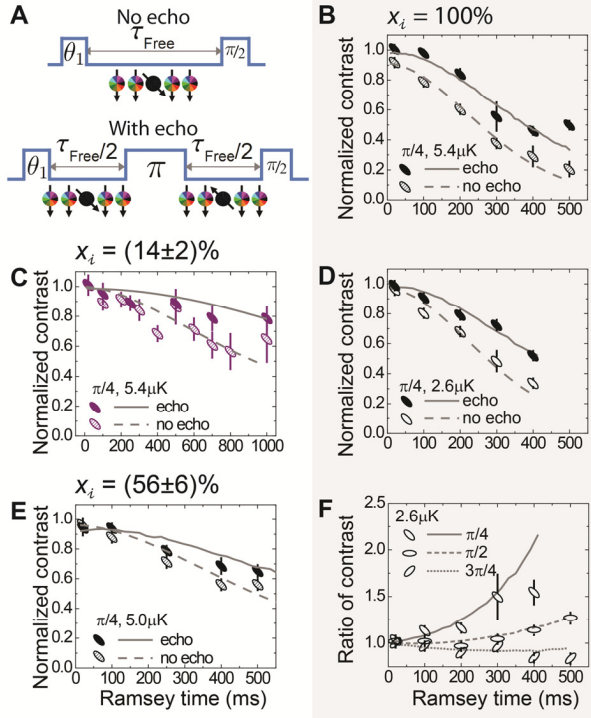
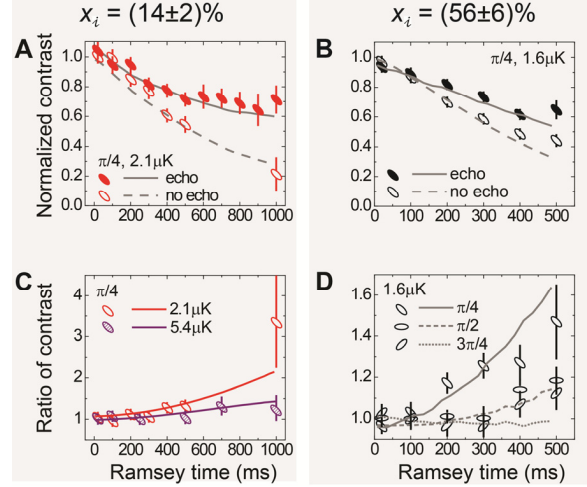

**Fig. 3. Evolution of orbital coherence under varying spin configurations.** (**A**) (Upper panel) Ramsey sequence with varying $\theta_1$ and $\tau_{\text{free}}$; (Lower panel) sequence with an echo ($\pi$) pulse. The group of circles illustrates the orbital configurations for interrogated atoms (black circles) and spectator atoms (colored circles). The contrast is normalized by comparing the high-atom-number raw Ramsey fringe contrast, $\mathcal{C}$ (defined in the main text), against the low-atom-number raw contrast. The filled symbols are for echo measurements and the empty symbols without echo. (**B** to **E**) Normalized contrasts for $\theta_1 = \pi/4$, with different $x_i$ and $T_R$: (**B** and **D**) $x_i = 100\%$, $T_R = 5.4$ μK and 2.6 μK, respectively; (**C**) $x_i = (14 \pm 2)\%$, $T_R = 5.4$ μK; (**E**) $x_i = (56 \pm 6)\%$, $T_R = 5.0$ μK. The solid and dashed lines show theory calculations with echo and without echo, respectively, using a two-orbital model where the spectators act as an effective magnetic field causing dephasing to the interrogated atoms. Under the conditions of (**B** to **E**), the dominant source for contrast decay arises from $p$-wave interactions between the interrogated atoms. (**F**) Effects of echo, characterized by the ratio of contrast with echo to that without echo, for $\theta_1 = \pi/4$ (anti-diagonal ellipse and solid line), $\pi/2$ (horizontal ellipse and short dashed line), and $3\pi/4$ (diagonal ellipse and short dotted line), under $x_i = 100\%$. Error bars represent 1σ standard error inflated by $\sqrt{\chi^2_{\text{reduced}}}$.

**Fig. 4. Evolution of orbital coherence with significant mode-changing processes.** (**A**) Normalized Ramsey contrast for $\theta_1 = \pi/4$, $x_i = (14 \pm 2)\%$, $T_R = 2.1$ μK: measurements with echo (filled symbols) and without echo (empty symbols); calculations with echo (solid line) and without echo (dashed line), respectively. (**B**) Normalized contrast for $\theta_1 = \pi/4$, $x_i = (56 \pm 6)\%$, $T_R = 1.6$ μK. (**C** and **D**) Effects of echo, characterized by the ratio of contrast with echo to that without echo, using the same symbol scheme as that in Fig. 3F. The solid, short-dashed and short-dotted lines are theory calculations that account for the contrast decay arising from the $p$-wave interactions between interrogated atoms, as well as mode-changing processes facilitated by $s$-wave interactions. The latter is accounted for as a time-dependent inhomogeneous dephasing not removable by a single echo. The theory uses a single fitting parameter for the mode-changing processes for all plots. Error bars represent 1σ standard error inflated by $\sqrt{\chi^2_{\text{reduced}}}$.

**Table 1.** *S*- and *P*-wave scattering lengths in Bohr radius ($a_0$).

*S*- and *P*-wave scattering lengths in Bohr radius ($a_0$)

| Channel | *S*-wave | *P*-wave | Determination |
|---|---|---|---|
| *gg* | 96.2(1) | 74.6(4) | [S-wave] Two-photon photo-associative spectroscopy [20] and ro-vibrational spectroscopy [21]<br>[P-wave] Analytic relation between S- and P-wave parameters [28] |
| *eg*$^+$ | 169(8) | -169(23) | [S-wave] Analytic relation [28]<br>[P-wave] Density shift in a polarized sample [17] |
| *eg*$^-$ | 68(22) | $-42^{+103}_{-22}$ | [S-wave] Density shift in a spin mixture at different temperatures (this work)<br>[P-wave] Analytic relation [28] |
| *ee* (elastic) | 176(11) | -119(18) | [S-wave] Analytic relation [28]<br>[P-wave] Density shift in a polarized sample [17] |
| *ee* (inelastic) | $\gamma_{ee} = 46^{+19}_{-32}$ | $\beta_{ee} = 121(13)$ | Two-body loss measurement [27] and analytic relation [28] |

## ACKNOWLEDGEMENTS

We thank P. Julienne, B. Gadway, T. Nicholson, B. Bloom, and A. Gorshkov for technical discussions. We acknowledge funding for this work from NIST, NSF PFC, AFOSR (MURI), AFOSR, DARPA QuASAR, and Austrian Science Foundation, SFB FoQus, ERC Synergy Grant UQUAM and SIQS. M.B. acknowledges support from the National Defense Science and Engineering Graduate fellowship program and the NSF Graduate Research Fellowship program.

After submission of this work, we became aware of an independent study of SU(*N*) physics in Yb atoms (F. Scazza *et al.*, arXiv:1403.4761).

Supplementary Material for

# Spectroscopic observation of SU(*N*)-symmetric interactions in Sr orbital magnetism

X. Zhang, M. Bishof, S. L. Bromley, C.V. Kraus, M. S. Safronova, P. Zoller,

A. M. Rey, J. Ye

**This PDF file include:**

Materials and Methods

Supplementary Text

Fig. S1

Tables. S1 to S3

References [31–46], following [1–30] in the main text

## Materials and Methods

*Preparation of atomic samples at given temperatures and spin configurations*

We trap 600 to 3000 $^{87}$Sr atoms in a one-dimensional (Z) optical lattice at 813 nm, and cool them to ∼2 $\mu$K along Z. Four types of nuclear spin configurations are generated: (1) a single spin state via optically pumping atoms into $m_I = +9/2$ with a circularly-polarized laser beam; (2) a two-spin mixture via pumping into $m_I = +9/2$ and $-9/2$ with a $\pi$-polarized beam; (3) a poorly polarized sample via weakening the circularly-polarized pumping laser power and applying a subsequent depolarizing beam; (4) an almost unpolarized sample by applying no optical pumping. To cool atoms along Z, we apply resolved sideband cooling for (1)∼(3) and Doppler cooling for (4). In the radial (X-Y) directions, atoms are cooled to about 2 $\mu$K with Doppler cooling and are at 5 ∼ 6 $\mu$K without cooling in lattice.

*Statistical methods for data analysis*

Following our previously reported procedure [6], the data are first binned into small chunks, the means and standard deviations of these bins are determined, and a reduced chi-squared, $\sqrt{\chi^2_{\text{reduced}}}$, is determined. If $\sqrt{\chi^2_{\text{reduced}}} > 1$, indicating that the data are over-scattered, we inflate the bins' standard deviations to bring $\sqrt{\chi^2_{\text{reduced}}}$ to 1, and the weighted average procedure is then performed again. This approach is employed to all measurements in this work.

*Differential measurements*

In order to determine the effect of inter-atomic interactions, we employ differential measurements throughout this work. In the density-dependent frequency shift measurements shown in Fig. 2, we measure the difference between transition frequencies at high and low atom numbers, where the atom number is modulated by a factor of $2 \sim 3$, normalize the frequency difference by the atom number difference, and then scale it to a given atom number (4000 interrogated atoms in Fig. 2).

In the Ramsey contrast decay measurements shown in Figs. 3 and 4, we remove the single-particle decoherence effect in a differential measurement protocol. We take measurements at alternate high and low atom numbers and then normalize the raw contrast measured at high atom numbers by that measured at the lowest workable atom number ($\sim 600$). The high-normalized-by-low contrast ratio is recorded as the normalized contrast, shown as the vertical axes in Figs. 3B to 3E and Figs. 4A,4B.

Here the raw Ramsey contrast is determined as follows: under a varying phase of the final $\pi/2$ pulse, we measure the Ramsey fringe contrast by either determining the variance of excitation fraction [17] for types (1) to (3) of nuclear spin configurations (see the first section on atomic sample preparation), or by directly fitting fringes with a sinusoidal function for type (4) spin configurations, where the interrogated fraction is small($\sim 10\%$), in order to reject technical noise.

**Two-orbital SU(*N*) spin model**

*Hamiltonian*

We consider an array of ultracold $^{87}$Sr atoms trapped in a 1D optical lattice at the magic wavelength (*i.e.* same trapping for $|g\rangle$ and $|e\rangle$). The lattice potential tightly confines the atoms in the lowest vibrational mode along the *Z*-direction and generates a weak harmonic confinement with angular frequency $\omega_R = 2\pi\nu_R$ along the radial directions.

We first consider the dynamics of $\mathcal{N}$ atoms in a single lattice site and label the thermally populated oscillator modes $\mathbf{n}_j = (n_{X_j}, n_{Y_j})$ with $j \in \{1, \ldots, \mathcal{N}\}$. At typical operation conditions of $\nu_R \sim 500 - 600$ Hz and temperatures in the $\mu$K regime, the mean interaction energy per particle is much weaker than the energy splitting between neighboring single-particle vibrational modes along any spatial direction. Thus, at leading order, only collisions between the atoms that conserve the total single particle energy must be considered. Such processes conserve the total number of particles per mode. In this case, the many-body dynamics is mainly governed by the internal degrees of freedom of the atoms, *i.e.* their electronic and nuclear spin degree of freedom, and the motional degrees can assumed to be frozen. If the gas is initially prepared in such a way that there is at most one atom per mode, the interaction dynamics of the system can be mapped to a lattice Hamiltonian with spin and orbital degree of freedom, where the vibrational modes $\mathbf{n}_j$ correspond to the lattice sites.

Note, however, that in the case of a pure harmonic spectrum, mode changing collisions are energetically allowed even under weak interactions due to (i) the linearity of the harmonic oscillator spectrum and (ii) the separability of the harmonic oscillator potential along the *X* and *Y* directions. Condition (i) allows two particles in modes $(n_X, n_Y)$ and $(m_X, m_Y)$ to collide and scatter into modes $(n_X + k, n_Y + k')$ and $(m_X - k, m_Y - k')$ without violating the energy conservation constraint. Condition (ii) allows the same two particles to scatter into modes $(n_X, m_Y)$ and $(m_X, n_Y)$. Those issues, in principle, can impose important limitations on the validity of the spin model in a harmonic trap. In practice, however, the trapping potential is not fully harmonic. It comes from the Gaussian beam profile of the lasers and is given by $V_R \approx -Ae^{-\frac{2R^2}{w_0^2}}$ with $w_0$ the beam waist. To leading order, the trapping potential is harmonic $V_R \sim \frac{m\omega_R^2}{2}R^2$, but for an atom in a mode $\{n_X, n_Y\}$, there are higher order corrections of the energy beyond leading order: $E_\mathbf{n} = \hbar\omega_R(n_X + n_Y + 1) + \Delta E_\mathbf{n}$, with $\Delta E_\mathbf{n} \sim \hbar\omega_R \left(2\frac{a_{ho}^R}{w_0}\right)^2 \left(3(n_X^2 + n_Y^2) + 4n_X n_Y + 5(n_X + n_Y + 1)\right)$. At typical operating conditions: $\nu_R = \frac{\omega_R}{2\pi} \sim 500 - 600\text{Hz}$, $w_0 \sim 30\mu\text{m}$, $T > 1\mu\text{K}$, and a mean occupation mode number $\bar{n}_{X,Y} > 50$, the difference of $\Delta E_\mathbf{n}$ for nearby modes is larger than $2\pi\hbar \times 10$ Hz which is not negligible compared to typical interaction energy scales $\sim$Hz. The first term in $\Delta E_\mathbf{n}$ thus prevents processes (i), while the second term, which breaks the separability of the potential, prevents processes (ii). Based on this argument we first restrict our analysis to only processes that conserve the number of particles per mode. For details, see Ref.[24]

We denote by $\hat{c}^\dagger_{\sigma m \mathbf{n}_j}$ the creation operator of a fermion in the mode $\mathbf{n}_j$, the electronic state $\sigma = e, g$ and with nuclear spin $m = 1, \ldots, N \leq 2I + 1$. $N$ is chosen by initial state preparation. The field operator creating an atom at position $\mathbf{R}$ and with quantum numbers $\sigma, m$ can then be written as $\hat{\Psi}^\dagger_{\sigma,m}(\mathbf{R}) = \phi_0^Z(Z) \sum_\mathbf{n} \hat{c}^\dagger_{\sigma m \mathbf{n}} \phi_{n_X}(X) \phi_{n_Y}(Y)$. The functions $\phi_0^Z$ and $\phi_{n_{X/Y}}$ are the longitudinal and transverse vibrational modes. We introduce the spin orbital operators

$$\vec{T}_j = \frac{1}{2} \sum_{\alpha,\beta,m} \hat{c}^\dagger_{\alpha m \mathbf{n}_j} \vec{\sigma}_{\alpha\beta} \hat{c}_{\beta m \mathbf{n}_j}, \tag{S1}$$

where $\vec{\sigma}$ is the vector of Pauli matrices acting on the $\{e, g\}$ basis (set $g = 1$ and $e = 2$), and further the nuclear-spin permutation operators

$$S_n^m(j) = \sum_\sigma \hat{c}^\dagger_{\sigma n \mathbf{n}_j} \hat{c}_{\sigma m \mathbf{n}_j}. \tag{S2}$$

Those operators satisfy the SU($N$) algebra $[S_n^m(i), S_p^q(j)] = \delta_{i,j}[\delta_{m,p} S_n^q(i) - \delta_{n,q} S_p^m(i)]$ and generate SU($N$) rotations of nuclear spins.

At $\mu$K temperatures, it is a valid assumption to consider only $s$- and $p$- wave scattering channels, each characterized by four elastic scattering parameters. We denote them by $a_\eta$ ($s$-wave scattering length) and $b_\eta^3$ ($p$-wave scattering volume), where $\eta = gg, ee, eg^+, eg^-$, is used to denote scattering of two atoms in the states $|gg\rangle, |ee\rangle$ and $|\pm\rangle = (|ge\rangle \pm |eg\rangle)\sqrt{2}$, respectively. Let us first assume that the spin wave function of the two atoms is symmetric. Then, due to

the fermionic statistics only atoms in the singlet state $|-\rangle$ can experience $s$-wave collisions, while atoms in the triplet states $|gg\rangle, |ee\rangle$ and $|+\rangle$ can only scatter via $p$-wave interaction. The interaction Hamiltonian for symmetric nuclear spin wave function is then of the form $H^+_{\mathbf{n}_j,\mathbf{n}_{j'}} = \sum_{\eta=ee,gg,+} V^+_\eta |\eta\rangle\langle\eta| + U^-_{eg}|-\rangle\langle-|$. The interaction matrix elements $V^\pm_\eta$ and $U^\pm_\eta$ are defined via the $s$- and $p$-wave matrix elements $P_{\mathbf{n}_j,\mathbf{n}_{j'}}$ and $S_{\mathbf{n}_j,\mathbf{n}_{j'}}$ that depend on the harmonic oscillator modes as

$$(V^\pm_\eta)_{\mathbf{n}_j,\mathbf{n}_{j'}} = b^3_\eta P_{\mathbf{n}_j,\mathbf{n}_{j'},\mathbf{n}_{j'},\mathbf{n}_j} \equiv b^3_\eta P_{\mathbf{n}_j,\mathbf{n}_{j'}}, \tag{S3}$$

$$(U^\pm_\eta)_{\mathbf{n}_j,\mathbf{n}_{j'}} = a_\eta S_{\mathbf{n}_j,\mathbf{n}_{j'},\mathbf{n}_{j'},\mathbf{n}_j} \equiv a_\eta S_{\mathbf{n}_j,\mathbf{n}_{j'}} \tag{S4}$$

The coefficients $S_{\mathbf{nn'n''n'''}}$ and $P_{\mathbf{nn'n''n'''}}$ characterize $s$- and $p$-wave matrix elements, respectively, which depend on the vibrationl modes. Explicitly,

$$S_{\mathbf{nn'n''n'''}} = \frac{4\sqrt{2\pi}\sqrt{\omega_Z\omega_R}}{a^R_{ho}} \left[ s(n_X, n'_X, n''_X, n'''_X) s(n_Y, n'_Y, n''_Y, n'''_Y) \right], \tag{S5}$$

$$P_{\mathbf{nn'n''n'''}} = \frac{6\sqrt{2\pi}\sqrt{\omega_Z\omega_R}}{(a^R_{ho})^3} \left[ s(n_X, n'_X, n''_X, n'''_X) p(n_Y, n'_Y, n''_Y, n'''_Y) + p(n_X, n'_X, n''_X, n'''_X) s(n_Y, n'_Y, n''_Y, n'''_Y) \right], \tag{S6}$$

$$s(n, n', n'', n''') \equiv \frac{\int d\xi e^{-2\xi^2} H_n(\xi) H_{n'}(\xi) H_{n''}(\xi) H_{n'''}(\xi) d\xi}{\pi\sqrt{2^{n+n'+n''+n'''} n! n'! n''! n'''!}},$$

$$p(n, n', n'', n''') = \frac{\int d\xi e^{-2\xi^2} \left[ \left(\frac{dH_n(\xi)}{d\xi}\right) H_{n'}(\xi) - H_n(\xi) \left(\frac{dH_{n'}(\xi)}{d\xi}\right) \right] \left[ \left(\frac{dH_{n''}(\xi)}{d\xi}\right) H_{n'''}(\xi) - H_{n''}(\xi) \left(\frac{dH_{n'''}(\xi)}{d\xi}\right) \right]}{\pi\sqrt{2^{n+n'+n''+n'''} n! n'! n''! n'''!}}.$$

Here $H_n(x)$ are Hermite polynomials. $a^R_{ho}$ is the the transverse harmonic oscillator length.

Conversely, if the nuclear spin wave function of the two atoms is antisymmetric, then the triplet states can scatter via $s$-wave interactions, while the singlet state scatters via the $p$-wave channel. Thus, for antisymmetric nuclear spin wave function, the Hamiltonian is given by $H^-_{\mathbf{n}_j,\mathbf{n}_{j'}} = \sum_{\eta=ee,gg,+} U^+_\eta |\eta\rangle\langle\eta| + V^-_{eg}|-\rangle\langle-|$. Now, with the help of the spin orbital operators and dropping constant terms, we can rewrite $H^\pm_{\mathbf{n}_j,\mathbf{n}_{j'}}$ as

$$\frac{H^\pm_{\mathbf{n}_j,\mathbf{n}_{j'}}}{2\hbar} = J^\pm_{\mathbf{n}_j,\mathbf{n}_{j'}} (\vec{T}_j \cdot \vec{T}_{j'}) + \chi^\pm_{\mathbf{n}_j,\mathbf{n}_{j'}} \hat{T}^z_j \hat{T}^z_{j'} + C^\pm_{\mathbf{n}_j,\mathbf{n}_{j'}} \frac{\hat{T}^z_j + \hat{T}^z_{j'}}{2} + K^\pm_{\mathbf{n}_j,\mathbf{n}_{j'}} \mathbb{I}_{\mathbf{n}_j,\mathbf{n}_{j'}}, \tag{S7}$$

Here, $\mathbb{I}_{\mathbf{n}_j,\mathbf{n}_{j'}}$ is the identity matrix acting on the $j$ and $j'$ atoms. $J^\pm_{\mathbf{n}_j,\mathbf{n}_{j'}} = \frac{(\zeta^+_{eg} - \Upsilon^-_{eg})_{\mathbf{n}_j,\mathbf{n}_{j'}}}{2}$, $C^\pm_{\mathbf{n}_j,\mathbf{n}_{j'}} = \frac{(\zeta_{ee} - \zeta_{gg})_{\mathbf{n}_j,\mathbf{n}_{j'}}}{2}$, $\chi^\pm_{\mathbf{n}_j,\mathbf{n}_{j'}} = \frac{(\zeta_{ee} + \zeta_{gg} - 2\zeta^+_{eg})_{\mathbf{n}_j,\mathbf{n}_{j'}}}{2}$ and $K^\pm_{\mathbf{n}_j,\mathbf{n}_{j'}} = \frac{(\zeta_{ee} + \zeta_{gg} + \zeta^+_{eg} + \Upsilon^-_{eg})_{\mathbf{n}_j,\mathbf{n}_{j'}}}{8}$ where $\Upsilon = U, \zeta = V$ for + and $\Upsilon = V, \zeta = U$ for $-$. To write down the generic Hamiltonian $H_{\mathbf{n}_j,\mathbf{n}_{j'}}$ of two colliding atoms we define the projection operators $\mathscr{P}^-_{\mathbf{n}_j,\mathbf{n}_{j'}}$ and $\mathscr{P}^+_{\mathbf{n}_j,\mathbf{n}_{j'}}$ that project on the nuclear spin singlet an triplet state. They can be written in terms of the nuclear spin permutation operators as $\mathscr{P}^{\pm}_{\mathbf{n}_j,\mathbf{n}_{j'}} = \frac{\mathbb{I}_{\mathbf{n}_j,\mathbf{n}_{j'}} \pm \sum_{n,m=1}^{N} S_n^m(j) S_m^n(j')}{2}$. Then,

$$H_{\mathbf{n}_j,\mathbf{n}_{j'}} = \mathscr{P}^-_{\mathbf{n}_j,\mathbf{n}_{j'}} H^-_{\mathbf{n}_j,\mathbf{n}_{j'}} + \mathscr{P}^+_{\mathbf{n}_j,\mathbf{n}_{j'}} H^+_{\mathbf{n}_j,\mathbf{n}_{j'}}. \tag{S8}$$

This spin-orbital Hamiltonian is the SU(N) generalization of the so-called Kugel-Komskii (KK) Hamiltonian obtained for $N = 2$[31]. The KK Hamiltonian is used in condensed matter physics to model, e.g., the Mott insulator transition in metal oxides with perovskite structure[9]. Since the interactions are only pairwise, the total Hamiltonian describing the interactions between $\mathcal{N}$ atoms populating the modes $\vec{\mathbf{n}} = \{\mathbf{n}_1, \mathbf{n}_2, \ldots, \mathbf{n}_{\mathcal{N}}\}$ is given by $H_{\vec{\mathbf{n}}}^{SO} = \frac{1}{2}\sum_{j \neq j'} H_{\mathbf{n}_j,\mathbf{n}_{j'}}$. Note that the Hamiltonian commutes with all the $S_n^p(j)$ generators, $[H_{\vec{\mathbf{n}}}^{SO}, S_n^p(j)] = 0$, and thus it is SU(N) invariant.

In the presence of a laser field that drives the $g \to e$ transitions, the atom-light Hamiltonian in the rotating frame of the laser is given by:

$$\hat{H}_{\vec{\mathbf{n}}}^L(j)/\hbar = -\delta \hat{T}_j^z - \sum_{m=1}^{N} \Omega_m(j) S_m^m(j) \hat{T}_j^y + B \sum_{m=1}^{N} I_m S_m^m(j) \left[(g_e - g_g)\hat{T}_j^z + \frac{g_e + g_g}{2}\right]. \tag{S9}$$

Here, $I_m = m - (N+1)/2$, $\delta = \omega_L - \omega_0$, where $\omega_L$ is the laser frequency and $\hbar\omega_0$ the energy splitting between ground and excited state in the absence of a magnetic field. The second term is proportional to $\Omega_m(j)$ the corresponding Rabi frequency between the states $|gm\rangle$ and $|em\rangle$. $\Omega_m(j)$ depends on the harmonic oscillator level of the atom $\mathbf{n}_j$ [24] if there is any component of the probing laser wave-vector along the transverse direction. In our current operating conditions the mode dependence of $\Omega_m(j)$ can be neglected. $\Omega_m(j)$ also depends on the corresponding dipole matrix element between the $g - e$ transition of the nuclear spin sublevel $m$, determined by corresponding Clebsch-Gordan coefficients. The last term describes the Zeeman splitting in the presence of an external magnetic field $B$. Here, $g_{e,g}$ are the Lande-factors for the two electronic states. For $^{87}$Sr atoms, $\Delta g = (g_e - g_g) \sim 109$ Hz/G, which allows for the spectroscopic resolution of the nuclear spin sublevels [1, 32].

At our operating densities only the $e-e$ channels exhibits measurable inelastic collisions. In the presence of those, the use of a master equation to correctly capture the dynamics is required. However, at the mean field level, the recycling terms in the master equation vanish and one can just use an effective Hamiltonian with complex parameters [24], *i.e.* the terms $V^{ee}_{\mathbf{n}_j,\mathbf{n}_{j'}}, U^{ee}_{\mathbf{n}_j,\mathbf{n}_{j'}}$ are replaced by $V^{ee}_{\mathbf{n}_j,\mathbf{n}_{j'}} - \frac{i}{2}\Gamma^{ee}_{\mathbf{n}_j,\mathbf{n}_{j'}}, U^{ee}_{\mathbf{n}_j,\mathbf{n}_{j'}} - \frac{i}{2}\Lambda^{ee}_{\mathbf{n}_j,\mathbf{n}_{j'}}$. $\Gamma^{ee}_{\mathbf{n}_j,\mathbf{n}_{j'}}$ and $\Lambda^{ee}_{\mathbf{n}_j,\mathbf{n}_{j'}}$ have the same dependence as $V^{ee}_{\mathbf{n}_j,\mathbf{n}_{j'}}$ and $U^{ee}_{\mathbf{n}_j,\mathbf{n}_{j'}}$ but with the real part of the corresponding interaction parameters replaced by the imaginary one ($a_{ee} \to \gamma_{ee}$ and $b_{ee} \to \beta_{ee}$).

*Mean-field Dynamics*

We first derive mean-field equations of motion, making the ansatz $\hat{\rho} = \bigotimes_j \hat{\rho}_j$, where $\hat{\rho}_j = \sum_{\alpha,\beta=e,g} \rho^{m,m'}_{\alpha,\beta}(j)|\alpha,m\rangle\langle\beta,m'|$ is the density matrix of an atom in the oscillator level $\mathbf{n}_j$. $\rho^{m,m}_{\alpha,\alpha}$ are the number of atoms in nuclear spin $m$ and electronic orbital state $\alpha$ and $\rho^{m,m}_{\alpha,\beta\neq\alpha}$ correspond to orbital coherences. Since initially a nuclear spin statistical mixture is prepared and no nuclear spin coherences develop during the dynamics, we set $\rho_{\alpha,\beta}^{m,n\neq m}$ to zero:

$$\frac{d}{dt}\rho_{gg}^{mm}(j) = \Omega^m(j)\text{Re}[\rho_{eg}^{mm}(j)] - 2\sum_{k=1}^{\mathcal{N}} J_{\mathbf{n}_j,\mathbf{n}_k}^+ \text{Im}\left[\rho_{eg}^{mm}(k)\rho_{ge}^{mm}(j)\right] - 2\sum_{n=1}^{N}\sum_{k=1}^{\mathcal{N}} \bar{J}_{\mathbf{n}_j,\mathbf{n}_k}(1-\delta_{n,m})\text{Im}\left[\rho_{eg}^{nn}(k)\rho_{ge}^{mm}(j)\right],$$
(S10)

$$\frac{d}{dt}\rho_{ee}^{mm}(j) = -\Omega^m(j)\text{Re}[\rho_{eg}^{mm}(j)] + 2\sum_{k=1}^{\mathcal{N}} J_{\mathbf{n}_j,\mathbf{n}_k}^+ \text{Im}\left[\rho_{eg}^{mm}(k)\rho_{ge}^{mm}(j)\right] + 2\sum_{n=1}^{N}\sum_{k=1}^{\mathcal{N}} \bar{J}_{\mathbf{n}_j,\mathbf{n}_k}(1-\delta_{mn})\text{Im}\left[\rho_{eg}^{nn}(k)\rho_{ge}^{mm}(j)\right],$$
$$- \frac{1}{2}\left(\sum_{n=1}^{N}\sum_{k=1}^{\mathcal{N}}(\Gamma_{\mathbf{n}_j,\mathbf{n}_k}^{ee} + \Lambda_{\mathbf{n}_j,\mathbf{n}_k}^{ee})\rho_{ee}^{mm}(j)\rho_{ee}^{nn}(k) + \sum_{k=1}^{\mathcal{N}}(\Gamma_{\mathbf{n}_j,\mathbf{n}_k}^{ee} - \Lambda_{\mathbf{n}_j,\mathbf{n}_k}^{ee})\rho_{ee}^{mm}(j)\rho_{ee}^{m,m}(k)\right),$$
(S11)

$$\frac{d}{dt}\rho_{eg}^{mm}(j) = -i\rho_{eg}^{mm}(j)(I_m B\Delta g - \delta) + \frac{1}{2}\Omega^m(j)[\rho_{ee}^{mm}(j) - \rho_{gg}^{mm}(j)]$$
$$- i\rho_{eg}^{mm}(j)\sum_{k=1}^{\mathcal{N}}\left[C_{\mathbf{n}_j,\mathbf{n}_k}^+[\rho_{ee}^{mm}(k) + \rho_{gg}^{mm}(k)] + (J_{\mathbf{n}_j,\mathbf{n}_k}^+ + \chi_{\mathbf{n}_j,\mathbf{n}_k}^+)[\rho_{ee}^{mm}(k) - \rho_{gg}^{mm}(k)]\right.$$
$$+ (\bar{\chi}_{\mathbf{n}_j,\mathbf{n}_k} + \bar{J}_{\mathbf{n}_j,\mathbf{n}_k})\sum_{n=1}^{N}(1-\delta_{nm})[\rho_{ee}^{nn}(k) - \rho_{gg}^{nn}(k)] + \bar{C}_{\mathbf{n}_j,\mathbf{n}_k}\sum_{n=1}^{N}(1-\delta_{nm})[\rho_{ee}^{nn}(k) + \rho_{gg}^{nn}(k)]\right]$$
$$- i[\rho_{ee}^{mm}(j) - \rho_{gg}^{mm}(j)]\sum_{k=1}^{N}\left[-J_{\mathbf{n}_j,\mathbf{n}_k}^+ \rho_{eg}^{mm}(k) + \bar{J}_{\mathbf{n}_j,\mathbf{n}_k}\sum_{n=1}^{N}(1-\delta_{n,m})\rho_{eg}^{nn}(k)\right]$$
$$- \frac{1}{4}\sum_{k=1}^{\mathcal{N}}\rho_{eg}^{mm}(j)\left(2\Gamma_{\mathbf{n}_j,\mathbf{n}_k}^{ee}\rho_{ee}^{mm}(k) + (\Gamma_{\mathbf{n}_j,\mathbf{n}_k}^{ee} + \Lambda_{\mathbf{n}_j,\mathbf{n}_k}^{ee})\sum_{n=1}^{N}(1-\delta_{mn})\rho_{ee}^{nn}(k)\right).$$
(S12)

Here we have used the notation $\bar{A} \equiv (A^+ + A^-)/2$. Also note $(\rho_{eg}^{mm}(j))^* = \rho_{ge}^{mm}(j)$.

*TWA Approximation*

To account for the development of quantum correlations during the dynamics we use the truncated Wigner Approximation (TWA)[24]. The TWA has proven to be a successful approach to incorporate the leading quantum corrections to the mean-field dynamics. To implement the TWA, one needs to solve the mean-field equations of motion supplemented by random initial conditions distributed according to the Wigner function. For a spin coherent state with Bloch vector length $T^m = \mathcal{N}_m/2$ ($\mathcal{N}_m$ is the number of atoms in nuclear spin state $m$) and pointing initially along $-\hat{z}$, the Wigner function is given by:

$$\wp_m(T_{0x}^m, T_{0y}^m, T_{0z}^m) = \left(\frac{2}{\pi \mathcal{N}_m}\right) \delta(T_{0z}^m + \mathcal{N}_m/2) e^{\left[-2\frac{(T_{0y}^m)^2 + (T_{0x}^m)^2}{\mathcal{N}_m}\right]}. \quad (S13)$$

This Wigner function has a transparent interpretation. For a Bloch vector pointing along the direction $-\hat{z}$, the transverse components fluctuate because of the uncertainty principle. Note $T_x^m = \text{Re}(\rho_{eg}^{mm})$, $T_y^m = \text{Im}(\rho_{eg}^{mm})$ and $T_z^m = (\rho_{ee}^{mm} - \rho_{gg}^{mm})/2$.

For a statistical mixture, the total Wigner distribution is just a product of the Wigner distributions of the nuclear spin components, $\wp = \prod_m \wp_m$. Quantum mechanical expectation values of the spin operators relevant for this work can be computed as $\langle \mathscr{O}(\tau) \rangle_{\text{TWA}} = \int \left[\left(\prod_{m, \alpha = x, y, z} dT_{0\alpha}^m\right) \wp[\{T_{0\alpha}^m\}] \langle \mathscr{O}(\tau) \rangle\right]$, with $\langle \mathscr{O}(\tau) \rangle$ the classical evolution of the observable calculated using the mean-field equations.

*Large magnetic field limit*

Consider the experimental situation in which a single nuclear spin $m_I = m_i = 1$ is interrogated. Denote $\mathcal{N}_i$ the number of atoms in the interrogated nuclear state, and $\mathcal{N}_S = \Sigma_{\alpha=2}^N \mathcal{N}_\alpha$ spectator atoms. In the presence of a large magnetic field, $B \Delta g \gg \bar{J} \mathcal{N}$, flip-flop processes become energetically costly and the mean-field equations of motion simplify. After rotating out the fast oscillatory terms, during the dark time they become:

$$\frac{d}{dt}\rho_{\alpha\beta}^{mm>1}(j) \approx 0 \quad (S14)$$

$$\frac{d}{dt}\rho_{gg}^{1,1}(j) \approx 0 \quad (S15)$$

$$\frac{d}{dt}\rho_{ee}^{1,1}(j) \approx -\rho_{ee}^{1,1}(j) \sum_{k=1}^{\mathcal{N}} \Gamma_{\mathbf{n}_j, \mathbf{n}_k}^{ee} \rho_{ee}^{1,1}(k), \quad (S16)$$

$$\frac{d}{dt}\rho_{eg}^{1,1}(j) \approx -i\rho_{e,g}^{1,1}(j)(I_1 B \Delta g - \delta) \quad (S17)$$

$$-i\rho_{e,g}^{1,1}(j) \sum_{k=1}^{\mathcal{N}} \left[ C_{\mathbf{n}_j, \mathbf{n}_k}^+ \left(\rho_{gg}^{1,1}(k) + \rho_{ee}^{1,1}(k)\right) + \chi_{\mathbf{n}_j, \mathbf{n}_k}^+ \left(\rho_{ee}^{1,1}(k) - \rho_{gg}^{1,1}(k)\right) - \frac{i}{2}\Gamma_{\mathbf{n}_j, \mathbf{n}_k}^{ee} \rho_{ee}^{11}(k) \right.$$

$$\left. + (\bar{C}_{\mathbf{n}_j, \mathbf{n}_k} - \bar{\chi}_{\mathbf{n}_j, \mathbf{n}_k} - \bar{J}_{\mathbf{n}_j, \mathbf{n}_k}) \sum_{n=1}^{N} (1 - \delta_{nm}) \rho_{gg}^{nn}(k) \right]. \quad (S18)$$

where we have used the fact that all the spectator atoms were initially in the ground state. From those equations it is clear that the spectator atoms remain in the ground state and act just as a magnetic field on the interrogated atoms. In this limit TWA approximation only generated many-body correlations on the interrogated atoms.

*Modeling the density shift*

To model the density shift, one can use the collective mode approximation and replace the coupling constants by their collective thermal averages: $O^{\pm}_{\mathbf{n}j,\mathbf{n}_k} \to \overline{O^{\pm}}$. Under this approximation and in the presence of a large magnetic field, from Eq. (S18) one gets that the density shift has two independent contributions,

$$\Delta \nu = \Delta \nu^i + \Delta \nu^S. \qquad (S19)$$

The first term comes from *p*-wave interactions between interrogated atoms themselves and is given by

$$\Delta \nu^i \sim \mathcal{N}_i \left( \overline{C^+} - \cos\theta_1 \overline{\chi^+} \right), \qquad (S20)$$

with $\overline{C^+} = \frac{[b_{ee}^3 - b_{gg}^3]}{2} \langle P \rangle_{T_R}$ and $\overline{\chi^+} = \frac{[b_{ee}^3 + b_{gg}^3 - 2b_{eg}^{+3}]}{2} \langle P \rangle_{T_R}$. Here $\langle S \rangle_{T_R}$ and $\langle P \rangle_{T_R}$ correspond to thermal averages of the *s*- and *p*-wave overlap matrix elements. Assuming a Boltzmann distribution of initially populated modes, they depend on $T_R$ as $\langle S \rangle_{T_R} \propto \frac{1}{T_R}$ (decreases with $T_R$) and $\langle P \rangle_{T_R} \propto T_R^0$ (insensitive to $T_R$) [17].

The second term, $\Delta \nu^S$, is generated by the spectator atoms which act as an effective magnetic field along z given by

$$\Delta \nu^S = \overline{\Lambda} \mathcal{N}_S, \qquad (S21)$$

with $\overline{\Lambda} = \frac{(a_{eg}^+ + a_{eg}^- - 2a_{gg})}{4} \langle S \rangle_{T_R} + \frac{[b_{eg}^{+3} + b_{eg}^{-3} - 2b_{gg}^3]}{4} \langle P \rangle_{T_R}$.

Experimentally, two-body *e-e* losses are taken into account during the dark time, $\tau_{\text{free}}$, by measuring the time-average population (extracted from independent measurements periodically inserted into the clock sequence)[17]. Here for the sake of simplicity in deriving Eq.(S20) we ignored two-body *e-e* losses. However, losses can be easily included by solving Eqs.(S17) and (S18) and then using for the density shift the computed time average population. In general losses can give rise to a small non-linear dependence of the density shift with excitation fraction, especially at high densities, but for the current experimental atom numbers and trapping conditions those give a correction of the slope less than 1%.

*Modeling the contrast decay*

While a mean-field treatment is good enough to model the density shift, to model the contrast decay we need to include corrections from the TWA. Those corrections account for the development of quantum correlations during the dynamics. We denote $\tilde{\rho}_{\sigma\sigma'}^{mm}$ the density matrix components obtained after performing the Truncated Wigner average as well as an average over the number and thermal distributions in the various pancakes.

In terms of the density matrix components, the Ramsey fringe contrast is given by $\mathscr{C}(\tau_{\text{free}}) = 2\left|\tilde{\rho}_{eg}^{11}(\tau_{\text{free}})\right| / \left[\tilde{\rho}_{ee}^{11}(\tau_{\text{free}}) + \tilde{\rho}_{gg}^{11}(\tau_{\text{free}})\right]$, where $\tau_{\text{free}}$ is the free evolution time as illustrated in Fig. 3A. In the presence of a large magnetic field, exchange is energetically suppressed and the ground state population of the various nuclear spin sublevels is conserved. In this limit, the decay of the Ramsey contrast $\mathscr{C}$ can be split into two separate contributions. One coming from the decay due to interactions between interrogated atoms themselves, $\mathscr{C}^i(\tau_{\text{free}})$, and another arising from the interactions between interrogated and spectator atoms, $\mathscr{C}^S(\tau_{\text{free}})$:

$$\mathscr{C}(\tau_{\text{free}}) = \mathscr{C}^i(\tau_{\text{free}})\mathscr{C}^S(\tau_{\text{free}}). \tag{S22}$$

$\mathscr{C}^i(\tau_{\text{free}})$ was measured and investigated in Refs.[17, 24] and has contributions coming from four different mechanisms: (i) At the single-site level (with fixed $\mathscr{N}_i$) a decay of coherences due to the development of genuine many-body correlations proportional to $e^{-\frac{\mathscr{N}_i-1}{2}\left(\sin\theta_1 \tau_{\text{free}}\overline{\chi^+}\right)^2}$. (ii) Dephasing caused when averaging over sites with different $\mathscr{N}_i$ due to the atom number dependent precession rate of the coherences (Eq.(S20)). (iii) Decay due to two-body *e-e* inelastic collisions. The latter is partially compensated by normalizing the coherences by the total number of interrogated atoms. (iv) Decay due to off-resonant virtual excitations of motional states, accounted for as higher order terms in the spin Hamiltonian. All of those effects are accounted for theoretically. In this work, we apply the same model and same *p*-wave scattering parameters as those extracted in Ref. (17).

In the large B limit, $\mathscr{C}^S(\tau_{\text{free}})$ can be understood as a dephasing mechanism induced by an effective inohmogeneous and density-dependent magnetic field along z generated by the spectator atoms on the interrogated atoms. Under the frozen mode approximation, for an interrogated atom in transverse mode $\mathbf{n}_j$, interacting with $\mathcal{N}_S$ spectator atoms, the effective magnetic field is $\mathscr{B}_j^F = \mathcal{N}_S \Lambda_j$, where $\Lambda_j = \frac{1}{\mathcal{N}_S}\Sigma_{k \in \mathcal{M}^S}\Lambda_{\mathbf{n}_j,\mathbf{n}_k}$ with $\Lambda_{\mathbf{n}_j,\mathbf{n}_k} = \frac{(a_{eg}^+ + a_{eg}^- - 2a_{gg})}{4}S_{\mathbf{n}_j,\mathbf{n}_k} + \frac{[(b_{eg}^+)^3 + (b_{eg}^-)^3 - 2(b_{gg})^3]}{4}P_{\mathbf{n}_j,\mathbf{n}_k}$. $\mathcal{M}^{S,i}$ are the manifolds of thermally populated transverse modes occupied by spectators ($S$) or interrogated ($i$) atoms in that site, respectively. Consequently,

$$\mathscr{C}_{\text{frozen}}^S(\tau_{\text{free}}) = \left|\left\langle \frac{1}{\mathcal{N}_i}\Sigma_{j=1}^{\mathcal{N}_i} e^{i\mathscr{B}_j^F \tau_{\text{free}}}\right\rangle_{\mathcal{N},T_R}\right|, \tag{S23}$$

where $\langle . \rangle_{\mathcal{N},T_R}$ means average over the thermal and atom number distribution across the pancake array [24].

We find, however, that off-resonant virtual excitations of motional states, do play a role in the dynamics and are an additional source of contrast decay. Those generate a decay not removable by an echo pulse. On the contrary, $\mathscr{C}_{\text{frozen}}^S(\tau_{\text{free}})$ can be removed by echo. To account for those processes, we use a phenomenological adhoc model based on Ref. (24), which nevertheless seems to reproduce fairly well the experimental observations as shown in Fig. 4. Correspondingly, the contribution to contrast decay is given by

$$\mathscr{C}_{\text{mobil}}^S(\tau_{\text{free}}) = \left|\left\langle \frac{1}{\mathcal{N}_i}\Sigma_{j=1}^{\mathcal{N}_i} e^{i\mathcal{N}_S \vartheta_j \tau_{\text{free}}}\right\rangle_{\mathcal{N},T_R}\right|, \tag{S24}$$

and

$$\vartheta_j = \frac{y}{(\mathcal{N}^{\text{cut}}\mathcal{N}_S\mathcal{N}_i)}\Sigma_{(p\in\mathcal{M}^S),(q\in\mathcal{M}^i),l}S_{\mathbf{n}_j,\mathbf{n}_p,\mathbf{n}_p,\mathbf{n}_l}S_{\mathbf{n}_j,\mathbf{n}_q,\mathbf{n}_q,\mathbf{n}_l}, \tag{S25}$$

where $l$ sums over $\mathcal{N}^{\text{cut}}$ initially unpopulated modes chosen to be close in energy to the initially populated ones, and selected assuming a Boltzmann distribution with radial temperature $T_R$. $S_{\mathbf{n}_j,\mathbf{n}_p,\mathbf{n}_p,\mathbf{n}_k}$ are *s*-save transition matrix elements [see Eq. ( S5)]. $\mathcal{N}^{\text{cut}}$ and $y$ are fitting parameters, which are set to be the same for all cases shown in Figs. 3 and 4. $\vartheta_j$ decreases faster than $T_R^{-1}$ (more like $T_R^{-4}$) and the contribution of mode changing processes to be contrast decay can be neglected for hot temperature cases ($T_R > 5\ \mu K$).

**Interaction parameters**

*Relationship between s-wave and p-wave interaction parameters*

A multichannel quantum defect theory [28] predicts that, for a single van-der-Waals potential, $\eta$, the complex scattering lengths for the *s*-wave, $A_\eta = a_\eta - i\gamma_\eta$, and the complex scattering volumes for the *p*-wave, $B_\eta^3 = b_\eta^3 - i\beta_\eta^3$, are related with the van-der-Waals length $\bar{a} = \frac{2\pi}{\Gamma(\frac{1}{4})^2}\left(\frac{2\mu C_6}{\hbar^2}\right)^{1/4}, \Gamma(\frac{1}{4}) \approx 3.626$:

$$\frac{A_\eta}{\bar{a}} = 1 + \left(\frac{B_\eta}{\bar{a}}\right)^3 \left[\left(\frac{B_\eta}{\bar{a}}\right)^3 + 2.128\right]^{-1}. \tag{S26}$$

We have computed the $C_6$ coefficients for the $\eta = gg, ee$, and $eg$ channels (see later sections) and found them to be 3107(30) a.u., 5360(200) a.u., and 3880(80) a.u., respectively, where 1 a.u. = 1 $E_h a_0^6$, with $E_h$ beging the Hartree energy and $a_0$ being the Bohr radius. Those together with Eq.(S26) allow us to relate $A_\eta$ and $B_\eta$. Using prior measurements done in a nuclear spin polarized sample [17], we can determine $(a_{gg}, b_{gg})$, $(a_{ee} - i\gamma_{ee}, b_{ee} - i\beta_{ee})$, $(a_{eg}^+, b_{eg}^+)$. Using those parameters and current density shift measurements (Fig. 2), we determine $(a_{eg}^-, b_{eg}^-)$ as explained below.

*Extracting the parameters from measuremnst*

We use our knowledge of the *p*-wave volumes extracted from density shift and contrast measurements for polarized samples [17] and Fig. 2B, together with density shift measurements in the presence of spectator atoms (Figs. 2C,D) and the relationship between *s* and *p*-wave scattering parameters [28], Eq. (S26), to determine all the $^{87}$Sr interaction parameters in the four $e - g$ channels.

Specifically, from the slope and zero-point crossing of the density shift in a polarized sample we determine $[(b_{eg}^+)^3 - (b_{gg})^3]\langle P \rangle_{T_R} = -1.65\text{s}^{-1}$ and $[(b_{ee})^3 - (b_{gg})^3] = 0.4[(b_{eg}^+)^3 - (b_{gg})^3]$. For our trapping conditions we estimate $a_0^3 \langle P \rangle_{T_R} = (3.35 \pm 1)10^{-7}\text{s}^{-1}$ and $a_0 \langle S \rangle_{T_R} = (0.08 \pm 0.02)(1\mu\text{K})/[T_R(\mu\text{K})]\text{s}^{-1}$. Those measurements, supplemented by excited state loss measurements [27] allow us to determine the following interaction parameters: $b_{eg}^+ \approx (-169 \pm 23)a_0$, $b_{ee} \approx (-119 \pm 18)a_0$ and $\beta_{ee} \approx (121 \pm 13)a_0$. Using Eq. (S26) we can determine the corresponding s-wave parameters: $a_{eg}^+ \approx (169 \pm 8)a_0$, $a_{ee} \approx (176 \pm 11)a_0$ and $\gamma_{ee} \approx (46^{+19}_{-32})a_0$.

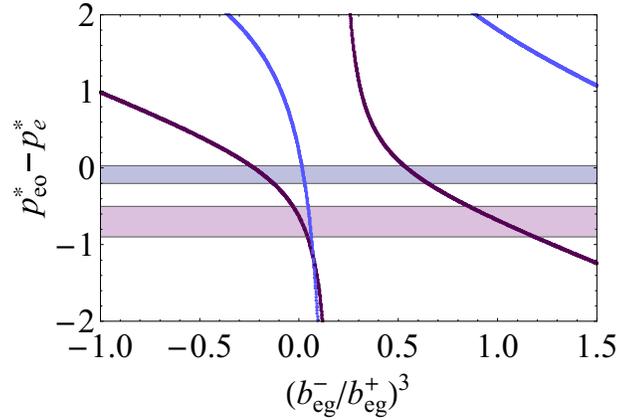

FIG. S1: **Fitting scattering parameters from the density shift data.** The figure shows the theoretical predictions of the offset of zero-shift excitation fraction $p_{e0}^* - p_e^*$ in a 13% spin mixture under two temperatures: 2.3(2) $\mu$K (blue line) and 6.5(4) $\mu$K (purple line) [computed using Eq.(S20-S21)] vs $(b_{eg}^-/b_{eg}^+)^3$. The shadowed regions correspond to the values of $p_{e0}^* - p_e^*$ that lie within the experimental error bars for the corresponding temperatures.

Finally from the density shift measurements in the presence of spectator atoms we determine $|b_{eg}^-| < 60a_0$ (best fit at $-42a_0$), and $a_{eg}^- \approx (68 \pm 22)a_0$. The large variation of $b_{eg}^-$ comes from the fact that the value that best fits the experimental parameters is in a region where it changes sign as shown in Fig. S1.

*Constraints on scattering parameters*

Taking into account all measurements, as well as the analytic relations betwen *s*-wave and *p*-wave scattering parameters, we establish 12 independent constraints for the 10 scattering parameters (8 elastic, 2 inelastic for *e-e*), see Table S1. Within experimental uncertainties, our results satisfy all these constraints and thus help determine the microscopic mechanism for interactions that obey SU(*N*) symmetry.

**Determination of $C_6$ coefficients of Sr: for the $^1S_0 + {}^1S_0$, $^3P_0 + {}^3P_0$, and $^1S_0 + {}^3P_0$ channels**

We investigate the molecular potentials asymptotically connecting to the $|A\rangle + |B\rangle$ atomic states. The wave function of such a system constructed from these states is

$$|M_A, M_B; \Omega\rangle = |A\rangle_{\text{I}} |B\rangle_{\text{II}}, \quad (S27)$$

where the index I(II) describes the wave function located on the center I(II) and $\Omega = M_A + M_B$. Here, the $M_{A(B)}$ is the projection of the appropriate total atomic angular momentum $\mathbf{J}_{A(B)}$ on the internuclear axis. We assume that $\Omega$ is a good quantum number for all calculations in this work (Hund's case (c)).

If *A* and *B* are the spherically symmetric atomic states and there are no downward transitions from either of them, the $C_6$ coefficient for the $A + B$ dimer is given by well known formula (see, e.g., Ref. [33])

$$C_6^{AB} = \frac{3}{\pi} \int_0^\infty \alpha^A(i\omega)\, \alpha^B(i\omega)\, d\omega, \quad (S28)$$

where $\alpha(i\omega)$ is the electric-dipole dynamic polarizability at an imaginary frequency. This formula is applicable to calculating $C_6$ coefficients for Sr-Sr $^1S_0 + {}^1S_0$, $^1S_0 + {}^3P_0$, and $^3P_0 + {}^3P_0$ dimers.

The integrals over $\omega$ needed for the evaluation of the $C_6$ coefficients are calculated using Gaussian quadrature of the integrand computed on a finite grid of discrete imaginary frequen-

TABLE S1: Constraints on scattering parameters

| Measurements or analytic relations | Constraints on scattering parameters | Number of independent constraints |
|---|---|---|
| Density shift for polarized atoms: slope and $p^*_{\text{pol}}$ (this work and Ref. 17) | $b_{gg}{}^3 + b_{ee}{}^3 - 2b_{eg}^+{}^3$ and $b_{ee}{}^3 - b_{gg}{}^3$ | 2 |
| Density shift for spin mixtures: $p^*$ (this work) | $b_{eg}^+{}^3 + b_{eg}^-{}^3 - 2b_{gg}{}^3$ and $a_{eg}^+ + a_{eg}^- - 2a_{gg}$ | 2 |
| Inelastic scattering [27] | $\beta_{ee}{}^3$ and $\gamma_{ee}$ | 2 |
| Measurements by other groups [20, 21] | $a_{gg}$ | 1 |
| Analytic relations between $s$- and $p$-wave parameters [28] | Relations for elastic scattering in all four channels: $gg, ee, eg^+, eg^-$ and inelastic scattering in the $ee$ channel | 5 |
| Total | | 12 |

cies [34, 35]. The integral in the expression for $C_6^{AB}$ coefficient is replaced by a finite sum

$$C_6^{AB} = \frac{3}{\pi} \sum_{k=1}^{N_g} W_k \, \alpha^A(i\omega_k) \, \alpha^B(i\omega_k) \tag{S29}$$

over values of $\alpha^A(i\omega_k)$ and $\alpha^B(i\omega_k)$ tabulated at certain frequencies $\omega_k$ yielding an $N_g$-point quadrature, where each term in the sum is weighted by the factor $W_k$. In this work, we use points and weights listed in Table A of Ref. [35] and $N_g = 50$.

Then, the calculation of the relevant $C_6$ coefficients is reduced to the calculation of $^1S_0$ and $^3P_0$ electric-dipole polarizabilities at imaginary frequencies. Such calculation for Yb has been discussed in detail in [36]. All calculations were carried out by two methods which allows us to estimate the accuracy of the calculations. The first method combines configuration interaction (CI) with many-body perturbation theory (MBPT) [37]. In the second method, which is more accurate, CI is combined with the coupled-cluster all-order approach (CI+all-order) that treats both core and valence correlation to all orders [38–40]. $C_6$ coefficients for Sr and Yb dimers and static electric-dipole $\alpha$ polarizabilities of Sr and Yb atoms [36, 41] are listed in Table S2. CI+all-order *ab initio* results are listed in rows labelled CI+all. The rows "Diff." give the relative size of the higher-order contributions estimated as the difference of the CI+all-order and CI+MBPT results. We note that the relative size of the higher-order contributions are different for Sr and Yb, due to different size of the higher-order corrections to dynamic polarizability at imaginary frequencies at large $\omega$. The $^3P_0 + ^3P_0$ Sr $C_6$ coefficient is compared with Refs. [42, 43].

The Sr-Sr and Yb-Yb $^1S_0 + ^3P_0$ $C_6$ coefficient can also be calculated using a semiempirical formula [44]

$$C_6^{AB} \approx \frac{2 \, \alpha^A(0) \, \alpha^B(0) \, C_6^{AA} \, C_6^{BB}}{C_6^{BB}(\alpha^A(0))^2 + C_6^{AA}(\alpha^B(0))^2}, \tag{S30}$$

where $\alpha^A(0)$ and $\alpha^B(0)$ are the electric-dipole static polarizabilities of the atomic states $A$ and $B$, and $C_6^{AA}$ and $C_6^{BB}$ are the $C_6$ coefficients for the $(A+A)$ and $(B+B)$ dimers, respectively.

While this formula gave the Yb $C_6(^1S_0 + ^3P_0)$ value in good agreement with the numerical CI+all-order result, it appears to be fortuitous. Most likely, the approximate formula worked well for Yb because of the similar contributions of high-orders to $^3P_0$ static polarizability and Yb $C_6(^3P_0+^3P_0)$ coefficient. The approximate formula result for $C_6(^1S_0+^3P_0)$ of Sr differs from our final number by 9%. Both Yb and Sr calculations are illustrated in Table S2.

TABLE S2: $C_6$ coefficients for Sr and Yb dimers in a.u. $\alpha$ is the electric-dipole static polarizability. CI+all-order *ab ibitio* results are listed in rows labelled "CI+all". The rows "Diff." give the relative size of the higher-order contributions estimated as the difference of the CI+all-order and CI+MBPT results. The uncertainties are given in parenthesis. The values listed in the column "Approx." are obtained using the approximate formula given by Eq. (S30). The values listed in the column "Numerical" are obtained using the formula given by Eq. (S28). [a]Adjusted by -1%, [b]Ref. [41]

|  |  | $\alpha(^1S_0)$ | $\alpha(^3P_0)$ | $C_6(^1S_0+^1S_0)$ | $C_6(^3P_0+^3P_0)$ | $C_6(^1S_0+^3P_0)$ | |
|---|---|---|---|---|---|---|---|
|  |  |  |  |  |  | Approx. | Numerical |
| Yb | CI+MBPT | 138 | 306 | 1901 | 3916 | 2492 | 2609 |
|  | CI+all [36] | 141(2) | 293(10) | 1929(39) | 3746(180) | 2487 | 2561(95)[a] |
|  | Diff. | 1.8% | -4.4% | 1.4% | -4.5% | -0.2% | -0.9% |
| Sr | CI+MBPT | 195.4 | 482.1 | 3091 | 5638 | 3517 | 3927 |
|  | CI+all | 197.8 | 458.1 | 3143 | 5553 | 3607 | 3958 |
|  | Diff. | 1.2% | -5.2% | 1.7% | -1.5% | 2.5% | 0.8% |
|  | **Best set** | 197.14[b] | 444.51[b] | **3107** | **5357** | **3548** | **3876** |
|  | **Final** |  |  |  | **5360(200)** |  | **3880(80)** |
|  | Ref. [42] |  |  |  | 5260(500) |  |  |
|  | Ref. [43] |  |  |  | 5102 |  |  |

To improve the accuracy of our Sr $C_6$ values, we extract the contributions of the dominant terms to the dynamic polarizabilities $\alpha(i\omega_k)$ in sum (S29) at each frequency and replace the corresponding energies by the experimental values and electric-dipole matrix elements by the recommended values that we determined in Ref. [41]. Then, Eq. (S29) is used with the modified dynamic polarizabilities to obtain the improved values of the $C_6$ coefficients. For the $5s^2\ ^1S_0$ dynamic polarizabilities, the contribution from the $5s^2\ ^1S_0 - 5s5p\ ^1P_1$ transition is replaced. For the $5s5p\ ^3P_0$ dynamic polarizabilities, the contributions from the $5s5p\ ^3P_0 - 5s4d\ ^3D_1$, $5s5p\ ^3P_0 - 5s6s\ ^3S_1$, $5s5p\ ^3P_0 - 5s5d\ ^3D_1$, and $5s5p\ ^3P_0 - 5p^2\ ^3P_1$ transitions are replaced. The best set energy and matrix element values are listed Table S3.

TABLE S3: CI+all-order (no small corrections) and final recommended matrix elements from Ref. [41] in a.u. The theoretical and experimental [45] transition energies are given in columns $\Delta E_{\text{th}}$ and $\Delta E_{\text{expt}}$.

| Transition | $\Delta E_{\text{th}}$ | $\Delta E_{\text{expt}}$ | $D_{\text{CI+all}}$ | $D_{\text{recom}}$ |
|---|---|---|---|---|
| $5s^2\ ^1S_0 - 5s5p\ ^1P_1^o$ | 21823 | 21698 | 5.272 | 5.248(2)[46] |
| $5s5p\ ^3P_0 - 5s4d\ ^3D_1$ | 3777 | 3842 | 2.712 | 2.675(13) |
| $5s5p\ ^3P_0 - 5s6s\ ^3S_1$ | 14673 | 14721 | 1.970 | 1.962(10) |
| $5s5p\ ^3P_0 - 5s5d\ ^3D_1$ | 20660 | 20689 | 2.460 | 2.450(24) |
| $5s5p\ ^3P_0 - 5p^2\ ^3P_1$ | 21208 | 21083 | 2.619 | 2.605(26) |

The resulting values are listed in the row "Best set" in Table S2 and are taken as final. The uncertainties of the $C_6(^3P_0 + ^3P_0)$ and $C_6(^1P_0 + ^3P_0)$ coefficients are estimated as the difference of the *ab initio* and the best set values.